\documentclass[aps,graphicx,reprint,twocolumn,groupedaddress]{revtex4-1} % for review and submission

\usepackage{graphicx}
\usepackage{dcolumn}
\usepackage{bm}
\usepackage{amssymb}

\begin{document}

\title{Point-Vortex Simulations Reveal Universality Class in Growth of 2D Turbulent Mixing Layers}
%\title{The Universal Growth Regime of 2D Turbulent Mixing Layers in Point-Vortex Temporal Simulations}

\author{Saikishan~Suryanarayanan}
 \email{saikishan.suryanarayanan@gmail.com.}
\author{Roddam~Narasimha}%
 \email{roddam@caos.iisc.ernet.in.}
\affiliation{Jawaharlal Nehru Centre for Advanced Scientific Research, Jakkur PO, Bangalore 560064, India}

\date{\today}

\begin{abstract}
A central but controversial issue in free turbulent shear flows has been the universality (or otherwise) of their growth rates. We resolve this issue here in the special case of a temporal 2D mixing layer in a point vortex gas by extensive high-precision numerical simulations, utilizing for the first time a powerful ensemble-averaging strategy. The simulations show that the momentum thickness of such a mixing layer grows at the universal asymptotic rate of $0.0167 (\pm 0.00017)$  times the velocity differential across the layer over a wide range of initial conditions, often after very long transients.
\end{abstract}

%\pacs{47.27.wj,47.27.Jv,47.32.C-}
\maketitle

The quest for universality in turbulent flow goes back to Reynolds \cite{Reynolds}, and includes the discovery of the log law in wall-bounded flows by Prandtl and Karman \cite{Schlichting} and the $k^{-5/3}$ spectrum by Kolmogorov \cite{Kolmogorov} . In turbulent shear flows, a unique `equilibrium' state independent of the detailed initial conditions has often been postulated \cite{Townsend}, but remains a controversial issue in wall-bounded \cite{Nagib,Marusic} as well as in free shear flows (e.g. wakes \cite{Narasimha72,Wygnanski86}, mixing layers \cite{Brown,Oster,Rogers,Balaras}).  The asymptotic nature of such postulated universalities, valid only at high Reynolds numbers ($Re \to \infty$)  and / or far downstream ($x \to \infty$), makes it difficult to be certain that the final state has been reached in flows with extremely long relaxation times \cite{Narasimha72,Sreenivasan}.

Mixing layers have been widely studied by experiments \cite{Brown,Oster,Mehta} , vortex simulations \cite{Acton,Delcourt,Aref80,Ghoneim} , DNS \cite{Rogers,Comte,Silvestrini,Sarkar,Wang} and LES \cite{Vreman,Balaras}. The dimensionless growth rate based on momentum thickness varies from 0.014 to 0.022 across experiments \cite{Rogers} and a similar scatter is observed among simulations as well. This has led to suggestions that there may be no universal growth rate independent of initial conditions \cite{Balaras,Oster}.

\begin{figure}
\includegraphics[width=3.5in]{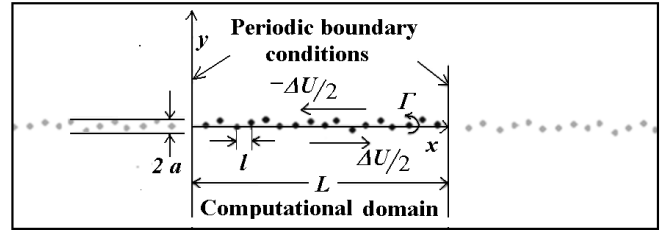}
\caption{\label{fig:schematic} Present point vortex model of the canonical temporal mixing layer, showing initial perturbation in the form of normal displacement of the vortices at $t = 0$ with amplitude measure $a$ .}
\end{figure}

We report here results of extensive high-precision  simulations of a 2D temporal turbulent mixing layer, for a wide class of initial conditions involving random and periodic normal displacements of a linear, equally spaced row of point vortices (separation distance $l$, Figure \ref{fig:schematic}) at the initial instant ($t = 0$). This can be seen as a direct `molecular dynamics' solution of a class of initial value problems we may pose in the spirit of the statistical mechanics of a point-vortex gas, formulated first by Onsager \cite{Onsager} (see \cite{Chavanis} for a recent review).  Although the flow considered is thus 2D instantaneously (and hence also in the mean), it is not irrelevant to a plane Navier-Stokes mixing layer (3D instantaneously but 2D in the mean). This is because experiments indicate that a 2D turbulence field provides a reasonable representation before the occurrence of the mixing transition \cite{Brown,Roshko,Konrad,Dimotakis}, and even later the coherent structures in the flow remain quasi-two-dimensional \cite{Wygnanski79}.

Point vortex simulations of a temporal mixing layer were pioneered by Delcourt and Brown \cite{Delcourt} and Aref and Siggia \cite{Aref80}, both of whom used  cloud-in-cell algorithms, and by Acton \cite{Acton}.  We revisit the problem using a different algorithm and the much more powerful computing resources now available, as the issue about universality still remains central to the subject.  The present simulations use double-precision 4th order Runge-Kutta integration to track each individual vortex, and (for the first time) provide averages over ensembles, involving upto $108$ realizations.  These substantial improvements in the numerics proved crucial for the conclusions we draw here.

We consider the limiting solution as $N \to \infty$, where $N \equiv L/l$ is the number of vortices in the  streamwise domain $L$ (Figure \ref{fig:schematic}).   With periodic boundary conditions imposed at $x  = 0, L$, the equations governing the motion of the vortices \cite{Acton} are

\begin{equation}
\frac{dx_i}{dt} = \sum_{j=1,j \neq i}^{N} \frac{ - \Gamma}{2L} \frac{\textrm{sinh}(2\pi(y_i-y_j)/L)}{\textrm{cosh}(2\pi(y_i-y_j)/L)-\textrm{cos}(2\pi(x_i-x_j)/L)}
\label{eq1}
\end{equation}

\begin{equation}
\frac{dy_i}{dt} = \sum_{j=1,j \neq i}^{N} \frac{\Gamma}{2L} \frac{\textrm{sin}(2\pi(x_i-x_j)/L)}{\textrm{cosh}(2\pi(y_i-y_j)/L)-\textrm{cos}(2\pi(x_i-x_j)/L)}
\label{eq2}
\end{equation}

\noindent where  $\Gamma$ is the (common) strength of each vortex and $(x_i,y_i)$ is the location of vortex $i$.

In the first instance the initial displacement of vortex $i$  in the array is drawn from a prescribed distribution of random numbers with zero mean and a measure of dispersion denoted by $a$ ; their subsequent motion is obtained by integrating (\ref{eq1} , \ref{eq2}). The solution for $\hat{\delta}$ , a measure of the thickness of the mixing layer at time $t$,  takes the form

\begin{equation}
\frac{\hat{\delta}}{l} = F_1 \left(\frac{t\Delta U}{l},\frac{L}{l},\frac{a}{l} \right)
\label{eq3}
\end{equation}

\noindent where $F_1$ is some (unprescribed) function.
Several exploratory simulations revealed an initial `transient' solution of (\ref{eq1} , \ref{eq2}) that is independent of $N$ at large $N$. If further this solution evolves to a state independent of the initial conditions for sufficiently large $t \Delta U / l$ , the dependence on  $a/l$  in (3) will disappear, yielding

\begin{equation}
\hat{\delta}/L = F_2 (t\Delta U/l) , t\Delta U/l >> 1
\label{eq4}
\end{equation}

\noindent where $F_2$ is an appropriate limit of $F_1$. In what may be called the `outer' limit, $t \Delta U/L = O(1)$, the solution may be expected to be of the form

\begin{equation}
\hat{\delta}/L = F_3 (t\Delta U/L)
\label{eq5}
\end{equation}

\noindent If we now postulate an overlap between solutions  (\ref{eq4}) and  (\ref{eq5}) the only possibility is that

\begin{equation}
\hat{\delta} = At\Delta U+B
\label{eq6}
\end{equation}

\noindent where $A$ and $B$ are independent of $t$. Equation  (\ref{eq6}) describes an `equilibrium' range in the problem. A universality class is defined here as a set of initial conditions $a/l$ for which $A$ is independent of $a/l$ in the simultaneous limits $t \Delta U/l \to \infty$ and $t \Delta U/L \to 0$, in the spirit of matched asymptotic expansions \cite{Dyke}.

The concept of (fluid-dynamical) equilibrium has been defined variously \cite{Townsend,Clauser}. We define equilibrium as a state in which the mean velocity field and the Reynolds shear stress both exhibit self-similarity with the same (time-dependent) length and velocity (alternatively time) scales.  (The Reynolds equation of momentum, see  (\ref{rans}) below, then implies (\ref{eq6}).)

In the present code, with the adopted time step of $0.1$ in $l/ \Delta U$, the distance a vortex moves during any time step does not exceed that to its nearest neighbor, and is almost always at least an order of magnitude less. A reduction in time step by a factor of $4$ did not materially affect the results. After every $100$ time steps the $x$ and $y$ components of velocity $(u, v)$ are computed on a grid of $20 L$ points in $x$ and $200$ points in $y$ using the Biot-Savart relation, and $x$-averaged quantities like the mean-velocity $ \overline{U}(y,t) = (1/L)\int_{0}^{L} dx [u(x,y,t)]$  are computed. The so-called momentum thickness, defined as \cite{Winant}

$ \theta(t) = \int_{-\infty}^{\infty} dy [0.25 - (\overline{U}(y,t)/\Delta U)^2]$

\noindent is often preferred here as a measure of layer thickness as the integral makes it more robust than the vorticity thickness and other point-based measures. However at $ t \Delta U/l < 10$,  $\theta$ can be misleading because of large overshoots in the mean velocity profile, and $\delta$ (defined as the separation in $y$ between the vortices at extreme $y$ positions) would be a more appropriate choice.(This explains its use in Regime I below.)

The accuracy of the algorithm has been assessed in several ways. The Hamiltonian, given for a cloud of\ $N$ point vortices by

$H = - (\Gamma^2 / 2\pi) \sum_{i<j}\ln (\mid \mathbf{r_i} - \mathbf{r_j} \mid/L) , \mathbf{r_i} \equiv (x_i,y_i)$, exhibits a maximum deviation (at $t \Delta U/l = 2500$) of $9\times10^{-6}$  of its initial value for $N = 3200$. The first moments of the vorticity distribution about the $x$- and $y$-axes  are conserved to within $10^{-16}$ and $3 \times 10^{-13}$ times $l$, and the second moment to within $1.3 \times 10^{-9}$ of its initial value. Another check is provided by mean momentum balance. For the present model the Reynolds-averaged Navier Stokes equation for $x$-momentum simplifies to

\begin{equation}
\frac{\partial \overline{U}}{\partial t} = -\frac{\partial ( \overline{u'v'})}{\partial y}  \equiv  \overline{v'\omega'} = \overline{v\omega}
\label{rans}
\end{equation}

\noindent where we have used the Reynolds decomposition $u = \overline{U} + u'$  etc., and $\omega = \overline{\omega}  + \omega'$ is the total vorticity (note that  $\overline{V} = 0$).  We find that the integral of (\ref{rans}), in what is defined as Regime II below, is satisfied to better than 0.05 \% (in simulations using 1600 vortices). These numbers show that the current computations are substantially more accurate than any previous work.

\begin{figure}
\includegraphics[width=3.5in]{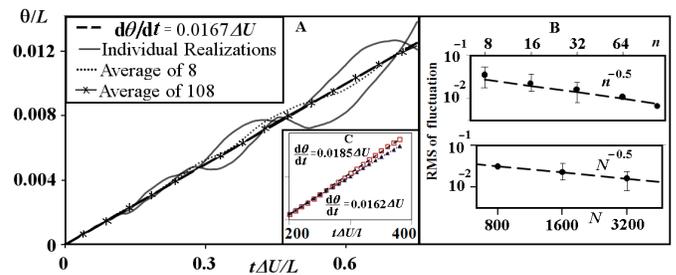}
\caption{\label{fig:Ensemble} A. Effect of ensemble averaging. B. Variation of RMS of relative departure of $\theta$ from (8) with $n$ $(N = 3200)$ and $N$ $(n = 32)$. C. Growth rates of two different ensembles($n = 10$) with same class of initial conditions and domain size; small ensembles of $10$ members and fits over narrow data ranges ($t_2/t_1 = 2$, where $t_1$ and $t_2$ are beginning and end of linear regime used to assess the growth rate) can result in large variations ($14\%$ in this case) in equilibrium growth rates. Note that $t_2/t_1$ and variation seen here are comparable to those presented by Balaras et al\cite{Balaras}}
\end{figure}

Finally, we note that averaging over a sufficiently large domain (analogous to long time averages in spatial mixing layer experiments) is essential to obtain accurate estimates of growth rate in temporal simulations. Averaging over a large ensemble of different realizations is an equivalent alternative (Figure \ref{fig:Ensemble}A). In the current simulations, the RMS fluctuation from the mean decreases with ensemble size $n$ like $n^{-\frac{1}{2}}$ (Figure \ref{fig:Ensemble}B), whereas the computational effort increases like $n$.  On the other hand the statistical fluctuations again decrease with number of vortices as $N^{-\frac{1}{2}}$, whereas the computational effort increases more rapidly like $N^2$.  Once $N$ is sufficiently large the former approach is thus computationally more economical, and is adopted here.

\begin{figure}
\includegraphics[width=3.5in]{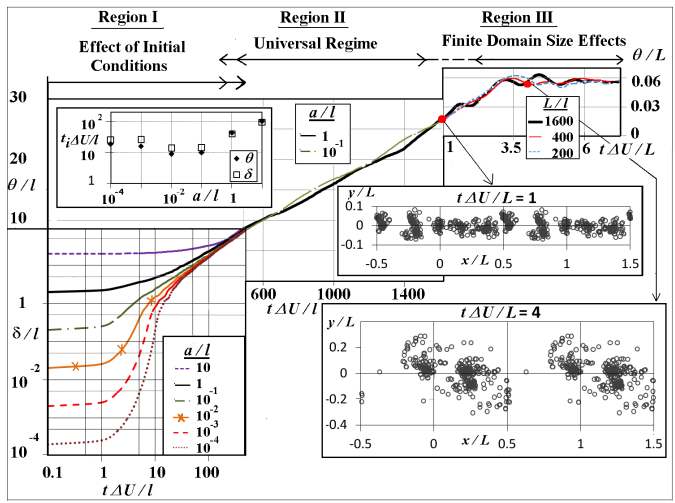}
\caption{\label{fig:Regimes} Composite diagram showing effect of initial conditions and domain size on the evolution of the mixing layer. Note use of $\delta$ and $\theta$ in different parts of the diagram, and change in the abscissa from $t \Delta U/l$ with a logarithmic scale upto 500, linear scale between $500$ and $1500$, and a switch to $t\Delta U/L$ thereafter.  Appropriate changes have been made on both abscissa and ordinate to ensure that the evolution curve should go smoothly from one regime to the next.  Inset on top left gives variation of initial transient with the amplitude of the initial vortex displacement.  Two insets on the right give pictures of the configuration of the vortices at $t\Delta U/L = 1$ (upper) and $4$ (lower).}
\end{figure}

Figure \ref{fig:Regimes} illustrates the evolution of the flow through results from selected simulations, with different domain sizes and initial conditions ($a/l$ ranging from $10^{-4}$ to $10$). It is seen that the variation of $\theta$ exhibits three distinct temporal regimes.

In Regime I, different initial conditions lead to widely different growth histories of $\delta$  from $t  = 0$ to a value of $t \Delta U / l$ that depends on  $a/l$. The duration $t_i$ of the transient (defined as the time at which the departure of $\delta$ or $\theta$ from the respective line of best fit in Regime II drops below 10\%) varies by an order of magnitude, and its dependence on  $a/l$ exhibits a minimum.  On the lower side of this minimum the disturbance is too small to quickly trigger turbulence, and on the other side it is too large to die down quickly. There are therefore optimal `trips' that lead to shortest transients.

In Regime II, which is between Regime I and $t \Delta U / L \sim 1$, growth is linear and independent of the initial conditions as well as domain size (this claim will be more elaborately supported below); in other words, the mixing layer is now in `equilibrium'.

In Regime III, beginning at $t \Delta U/L \sim 1$, the effects of finite domain size become noticeable and the scaling length changes over to $L$. This is due to the small number of coherent structures governing the dynamics in this regime (see insets in Figure \ref{fig:Regimes}). At lower $t \Delta U/L$ the effect is characterized by larger statistical fluctuations. These can in principle be reduced by averaging over a larger ensemble, but at $t \Delta U/L \sim 4$ there is only one structure left in the domain, with no further opportunity to amalgamate with others and grow.  Instead the structure just rotates about its own axis , resulting in oscillatory $\theta$  with a stationary mean.  A rough spatial analog of this effect of finite domain size is found in flow experiments carried out in smaller wind tunnels \cite{Narasimha73}.

\begin{figure}
\includegraphics[width=3.5in]{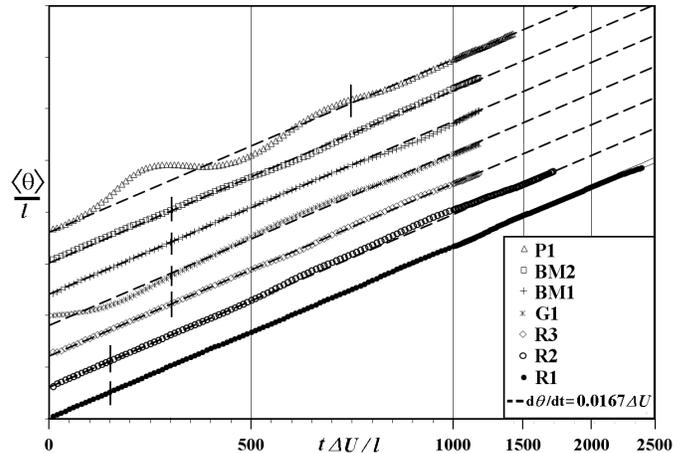}
\caption{\label{fig:Universality} Growth of momentum thickness for different simulations, shown with $y$-axis displaced. The universal equilibrium rate (\ref{universaleqn}) is shown in dashed lines, and is obtained by best fit to the average of 108 simulations with   $N  = 3200,  a/l = 0.05$ (R1).Note change in scales on both axes, each by the same factor, for $t\Delta U/l > 1000$. For tick marks see footnote on Table 1. R1 is shown with two thin lines within $\pm 1 \%$ of (\ref{universaleqn})}
\end{figure}

We now return to more detailed results on equilibrium linear growth in Regime II.  In order to test whether the growth rate is universal, a large set of simulations with widely different initial conditions and domain sizes have been performed (with each case averaged over a large ensemble : details are in Table \ref{tab:table1}). The initial vortex displacement distributions include the bi-modal type, in the form of sums of symmetric and asymmetric displaced Gaussians (respectively BM1, BM2).  The respective growth histories in Regime II are shown in Figure \ref{fig:Universality}. We take as reference, the best fit value for the largest ensemble simulated here, R1 ($n  = 108$),

\begin{equation}
\frac{d \theta}{dt} = 0.0167 \Delta U
\label{universaleqn}
\end{equation}

\noindent which extends over more than a decade in $t\Delta U/l$. RMS deviations from (\ref{universaleqn}) are listed in Table \ref{tab:table1}.

Periodic excitation needs a separate, detailed discussion, so we include here only one case to show that, after mimicking experimentally observed behaviour \cite{Oster} when wavelength is chosen as the relevant length scale, the layer eventually grows at the same universal rate (\ref{universaleqn}). The time taken to attain universality corresponds to over twice the length of the experimental test section in the 30 Hz case in Figure 13 of  Ref. \cite{Oster}.  It can be concluded from Figure \ref{fig:Universality} that there is a regime of linear growth, with a growth rate that is universal to within 1\%, for the wide class of initial conditions considered here.

\begin{table}
\caption{\label{tab:table1}}
\begin{ruledtabular}
\begin{tabular}{cccccc}
Code & $N$ & $n$ & Initial  & Best fit\footnotemark[1]   & RMS of \footnotemark[1] \footnotemark[2]    \\

&&&conditions&$ \frac{ d \theta}{d(t \Delta U)}$&  relative deviation  \\
&&&&&  from (\ref{universaleqn})\\

\hline

R1 & 3200 & 108 &	uniform random
 &	0.01667 &	0.00676 \\

&&& $a/l = 5 \times 10^{-2}$  &&\\
\hline
R2 &	10000	& 11 &	uniform random
 &	0.01662 &	0.02260 \\

&&& $a/l = 1 \times 10^{-1}$  &&\\
\hline

R3 &	1600 &	64 &	uniform random
 & 0.01666 &	0.00835 \\
&&& $a/l = 10^{-8}$	 && \\
\hline

G1 &	1600 &	64 &	Gaussian\footnotemark[3] & 0.01676 & 0.01845 \\
&&&  $ \sigma/ l = 5$	 &&\\
\hline

BM1 & 1600 &	64 &	Bi-modal\footnotemark[4] &	0.01654 &	0.01282\\
&&& $ \sigma_1 / l = 1 \times 10^{-1}$	 &&\\
&&& $\sigma_2 / l = 1 \times 10^{-1}$	 &&\\
&&& $d / l = 6 \times 10^{-1}$	 &&\\
\hline

BM2	& 1600 &	64 &	Bi-modal\footnotemark[4]
 &	0.01665&	0.01367\\

&&& $ \sigma_1 / l = 1 \times 10^{-2}$	 &&\\
&&& $\sigma_2 / l = 2 \times 10^{-2}$	 &&\\
&&& $d / l =  4 \times 10^{-2}$	 &&\\
\hline

P1 &	3200 &	32 &	Periodic forcing\footnotemark[5]
& 0.01667 &	0.00697\\

&&& $a/l = 4 \times 10^{-1}$ &&\\
&&& $\lambda/l = 100$	&& \\

\end{tabular}

\footnotetext[1]{Based on data beginning from the tick mark on each simulation in Figure 4 till the end of respective simulation.}
\footnotetext[2]{Defined as minimum with respect to $B$ of $\sqrt{\frac{1}{m} \sum_m{\left(\frac{\theta-(0.0167 t\Delta U + B)}{0.0167 t\Delta U + B}\right)^2}}, m$ - no. of data points}
\footnotetext[3]{$\sigma$ - Standard deviation.}
\footnotetext[4]{$\sigma_1$, $\sigma_2$ - Standard deviations of two Gaussians separated by $d$.}
\footnotetext[5]{0.1\% uniform random noise added to generate different realizations; $\lambda$ - Wavelength of periodic forcing}

\end{ruledtabular}
\end{table}

The present simulations yield growth rates of the same order as in experiments. This implies that the dominant mechanism in the growth of the momentum thickness in the mixing layer must be just the kinematics of the Biot-Savart relation, which includes the emergence of chaos even in a few-vortex system \cite{Aref83}. Scepticism about universality in the growth of (real) mixing layers \cite{Oster} stems from their known sensitivity to various factors, but an important additional factor suggested by the present simulations is the inadequacy of flow-development length in laboratory set-ups. Such a length is the spatial analogue of the long transient noted in some of the present simulations.  Similarly, as seen in Figure \ref{fig:Ensemble}C, short computational domains, fits made over small values of $t_2/t_1$ and small ensembles account for the observed variation in growth rates.Evidence for universality in the present simulations would be weak unless $L/l$ is in excess of $10^3$, averages are struck over sufficiently large ensembles and the equilibrium regime is sufficiently long   (Figure \ref{fig:Ensemble}). Similar factors can account for claims against universality in other types of simulation \cite{Balaras}. It is anyway of considerable interest that there exists at least one prototypical turbulent shear flow with three distinct regimes including one corresponding to a universal equilibrium state.

We are grateful to Drs. S.D. Sherlekar and R.K. Lagu for providing supercomputing resources at the Tata EKA and to Mr. Sapre for assistance in parallelizing our code at the Computational Research Laboratories, Pune.  We thank Dr. Garry Brown (Princeton) and Dr. Anatol Roshko (Caltech) for many rewarding and enjoyable discussions and suggestions, and Dr. Santosh Ansumali (JNCASR) for asking us to try bi-modal initial conditions.   We acknowledge support from DRDO through the project RN/DRDO/4124.

\end{document}